\documentclass[useAMS,usenatbib]{mn2e}
\usepackage{graphicx}

\title[The `Carina Flare' Supershell: Probing the Atomic and Molecular ISM in a Galactic Chimney\\]{The `Carina Flare' Supershell: Probing the Atomic and Molecular ISM in a Galactic Chimney\\}
\author[J. R. Dawson, N. Mizuno, T. Ohnishi, N. M. McClure-Griffiths and Y. Fukui]{J. R. Dawson$^{1}$\thanks{E-mail: joanne@a.phys.nagoya-u.ac.jp}, N. Mizuno$^{1}$, T. Onishi$^{1}$, N. M. McClure-Griffiths$^{2}$, and Y. Fukui$^{1}$\\
$^{1}$Department of Physics and Astrophysics, Nagoya University, Chikusa-ku, Nagoya, Japan\\
$^{2}$Australia Telescope National Facility, CSIRO, P.O. Box 76, Epping NSW 1710, Australia}
\begin{document}

\date{}

\pagerange{\pageref{}--\pageref{}} \pubyear{}

\maketitle


\begin{abstract}
The `Carina Flare' supershell, GSH 287+04-17, is a molecular supershell originally discovered in $^{12}$CO(J=1-0) with the NANTEN 4m telescope. We present the first study of the shell's atomic ISM, using H{\sc i} 21 cm line data from the Parkes 64m telescope Southern Galactic Plane Survey.
The data reveal a gently expanding, $\sim230\times360$ pc H{\sc i} supershell that shows strong evidence of Galactic Plane blowout, with a break in its main body at $z\sim280$ pc and a capped high-latitude extension reaching $z\sim450$ pc. The molecular clouds form co-moving parts of the atomic shell, and the morphology of the two phases reflects the supershell's influence on the structure of the ISM. We also report the first discovery of an ionised component of the supershell, in the form of delicate, streamer-like filaments aligned with the proposed direction of blowout. The distance estimate to the shell is re-examined, and we find strong evidence to support the original suggestion that it is located in the Carina Arm at a distance of $2.6\pm0.4$ kpc. Associated H{\sc i} and H$_2$ masses are estimated as $M_\mathrm{HI}\approx7\pm3\times10^5~\mathrm{M}_{\odot}$ and $M_{\mathrm{H}_2}\approx2.0\pm0.6\times10^5~\mathrm{M}_{\odot}$, and the kinetic energy of the expanding shell as $E_K\sim1\times10^{51}$ erg. We examine the results of analytical and numerical models to estimate a required formation energy of several $10^{51}$ to $\sim10^{52}$ erg, and an age of $\sim10^7$ yr. This age is compatible with molecular cloud formation time-scales, and we briefly consider the viability of a supershell-triggered origin for the molecular component.\\
\end{abstract}


\section{Introduction}

Supershells are some of the largest and most energetic structures in the Galactic ISM and play an important role in its evolution; disrupting and reshaping the gas on scales of hundreds of parsecs. As part of a multi-phase ISM, they are visible in many wavelength regimes; however, in the Milky Way they have been most extensively studied in the H{\sc i} 21 cm line. H{\sc i} is a powerful probe of the structure and dynamics of the atomic medium, and an excellent tracer of the large-scale morphology of neutral shells, which remain detectable even after their central energy sources have switched off. Beginning with the work of \citet{heiles79} almost 30 years ago, H{\sc i} surveys have lead to the detection of hundreds of Galactic shells and shell candidates \citep[e.g.][]{heiles79, heiles84, ehlerova05, mcclg02}.\par 

Most supershells are thought to be formed by stellar feedback from OB clusters. The stellar winds and supernovae inject vast amounts of energy ($10^{51}\sim10^{53}$ergs) into their surroundings, driving a shock front into the ISM and creating a giant cavity of hot, tenuous gas surrounded by a cool shell of swept-up neutral material \citep{bruh80, tomisaka81, mccray87}. Those with sufficient energy may expand rapidly along the vertical density gradient and break out of the Galactic disc, venting their hot interior gas into the Halo \citep{maclow88, tomisaka86}. Such Galactic `chimneys' are thought to play a central role in the disc--halo interaction, supplying the Halo with energy and metal-enriched material \citep{norman89}. However, at present only a handful of chimneys have been identified in the Milky Way (e.g. \citealp[GS 018-04+44,][]{callaway00}; \citealp[GSH 277+00+36,][]{mcclg03}; \citealp[GSH 242-03+37,][]{mcclg06}; \citealp[the W4 chimney,][]{ndeau96}), and there are few observational constraints on their characteristics.\par

Important questions also remain regarding the impact of supershells on the chemical and physical evolution of the disc ISM. In particular, their role in the manufacture, distribution and evolution of the molecular phase is still largely unknown. The accumulation of the ISM in such superstructures 
provides one means of generating the cool, dense conditions that favour molecule formation, and it has long been suspected that supershells may trigger the formation of molecular clouds \citep{bergin04, elmegreen98, hartmann01, mccray87}.
However, the mechanisms of this process and its relative importance in the Galactic ecosystem are still poorly understood. This has partly been due to a shortage of well-resolved observational candidates -- a problem that modern CO line surveys are proving able to address \citep{matsunaga01, moriguchi02, yamaguchi99}.

The `Carina Flare' \citep[hereafter F99]{fukui99} is an outstanding example of a molecule-rich supershell. Located in the Carina Arm at an estimated distance of $2.6\pm0.4$ kpc, it was first discovered with the NANTEN 4m telescope, which mapped the molecular ISM to a resolution of $2.6$ arcmin ($\sim2$ pc) in the 115 GHz $^{12}$CO(J=1-0) line. The molecular component is seen as a prominent, expanding cloud complex that extends unusually far above the Galactic plane, reaching $z\sim450$ pc at its extreme, 
and evidence of an atomic counterpart was provided by loop-like dust structure and an associated H{\sc i} cavity. Unlike the sparse molecular clouds, this atomic component should be an unparalleled tracer of the supershell's morphology; shedding light on its structure, dynamics and relationship to its local environment. However, previously available low resolution H{\sc i} data \citep{kerr86} reveals little more than the presence of an asymmetric cavity, and to date the atomic ISM has received only a cursory examination. 

This paper utilises H{\sc i} data from the Parkes 64m dish Southern Galactic Plane Survey \citep[SGPS,][]{mcclg05}, to present the first study of the atomic component of the Carina Flare supershell. With spatial and velocity resolutions of $\sim16$ arcmin ($\sim10$ pc at $2.6$ kpc) and $0.82$ km s$^{-1}$, the SGPS is a significant improvement over previous large-scale surveys of Galactic H{\sc i}. The data newly reveal the supershell's morphology and velocity structure, including signatures of Galactic Plane blowout, providing compelling evidence that the Carina Flare supershell may represent a new example of a young chimney system. They also reveal a rich variety of substructure, and intriguing morphological correspondences between the H{\sc i} and H$_2$ components, setting the stage for future detailed studies of the shell ISM.\par

This paper is conceived as the first of a series examining the Carina Flare supershell. Forthcoming work will present high-resolution comparisons between the molecular and atomic ISM, as well as investigations of the ionised component and stellar population, and will seek to explore the origin of the molecular component in detail.
The present work focusses on the global characteristics of the supershell and its relationship to the Galactic environment. \S 2 gives details of the Parkes SGPS H{\sc i} and NANTEN $^{12}$CO(J=1-0) observations. We present the datasets in \S 3, drawing attention to important morphological features and relationships, and estimating the mass of the neutral shell. We also compare with H$\alpha$ survey data, newly identifying an ionised component. \S 4 explores aspects of the supershell's evolution and relationship to the Galactic environment, presenting evidence for Galactic Plane blowout, discussing distance and location, and confirming the $2.6\pm0.4$ kpc distance estimate. We also discuss improved constraints on formation energy and age, and touch on the complex issue of triggered molecular cloud formation, before summarising in \S 5. A short appendix briefly outlines a simple method of fitting to the supershell in the $l$-$v_{lsr}$ plane. \par
For consistency with standard nomenclature, the Carina Flare supershell has been assigned the name GSH 287+04-17. 

\section{Observations}

\begin{figure}
\includegraphics{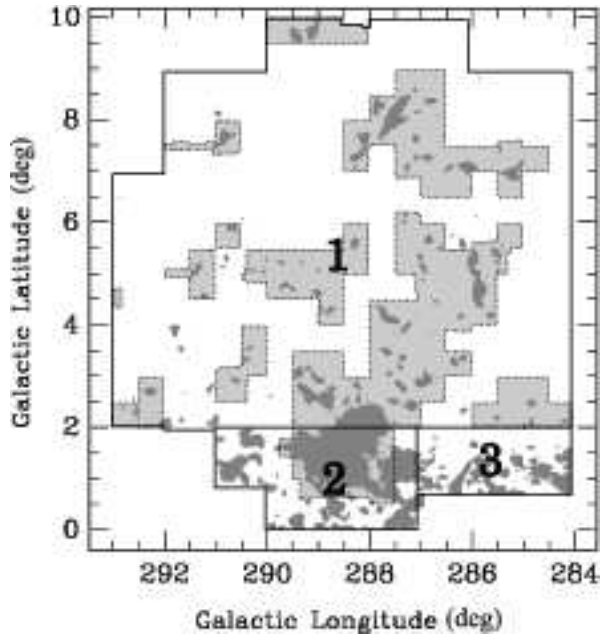}
\caption{Regions observed in $^{12}$CO(J=1-0) with the NANTEN telescope. The region numbers are referred to in the text. Light grey areas enclosed in dotted lines are observed at 2 arcmin grid spacing, and all other areas at 4 arcmin. Dark grey regions mark emission detected at the 3$\sigma$ level, corresponding to integrated intensities of $>1.5$K km s$^{-1}$, where the integration is over the velocity range $-35 < v_{lsr} < -3$ km s$^{-1}$.}
\label{obsregions}
\end{figure}

\begin{figure*}
\includegraphics{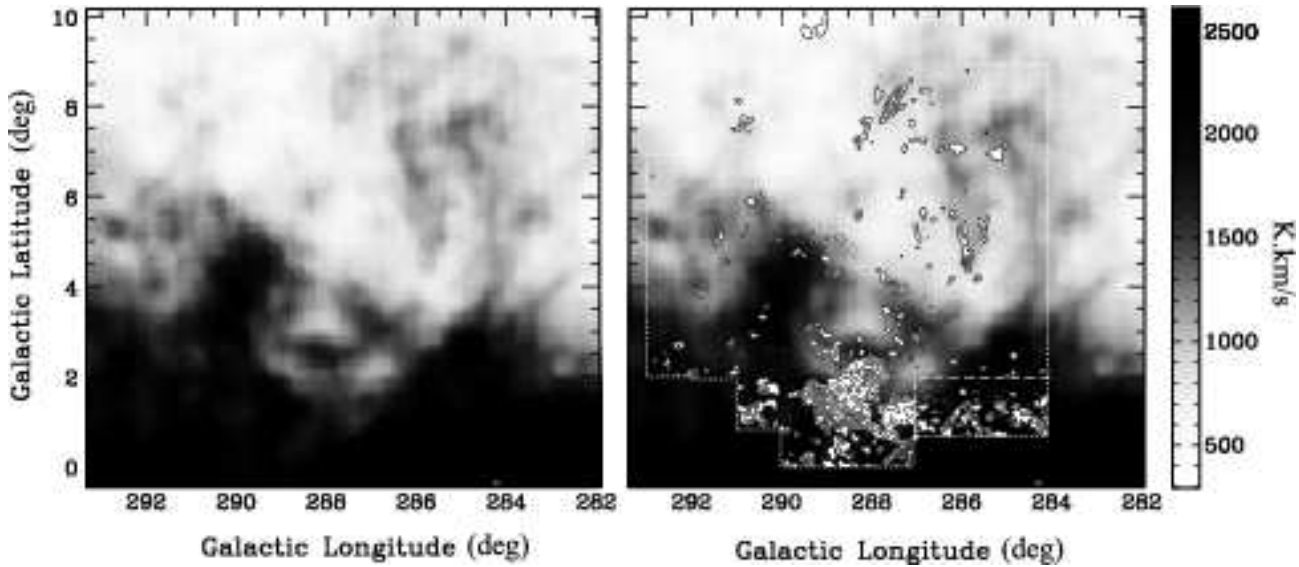}
\caption{Integrated intensity map of the Carina Flare supershell. Left: Parkes SGPS H{\sc i} 21 cm line data (grayscale). Right: overlaid with NANTEN $^{12}$CO(J=1-0) molecular clouds (filled contours). Both data sets are integrated over the velocity range $-36 < v_{lsr} < -3$ km s$^{-1}$. The lowest $^{12}$CO contour level is 1.5 K km s$^{-1}$ and the interval is 5 K km s$^{-1}$. The NANTEN observed regions are marked with dotted lines. Areas of noisy NANTEN data are enclosed in dashed lines and may contain some spurious emission at this cut-off level.}
\label{COHIii}
\end{figure*}

\subsection{Parkes 64m SGPS H{\sc i} Data}

The Southern Galactic Plane Survey mapped the 4th quadrant and part of the 1st quadrant of the Galactic plane in the 21 cm neutral hydrogen hyperfine structure line. The survey combined single dish observations from the Parkes 64m radio telescope and interferometric data from the Australia Telescope Compact Array (ATCA), to produce high resolution maps of the $-1.5 < b < 1.5^\circ$ region and low resolution maps extending to $b=\pm10^\circ$. The present paper makes use of the low resolution portion of the SGPS. The data were taken over a period of several months from June 1999 to March 2000, using an on-the-fly mapping technique. The effective resolution is $\sim$16 arcmin, re-gridded to a pixel size of 4 arcmin, and the velocity channel width is 0.82 km s$^{-1}$. The noise per channel is $\sim180$ mK, which is orders of magnitude lower than the typical signal. Full details of the observing strategy and data reduction can be found in McClure-Griffiths et al.'s (2005) explanatory paper.

\subsection{NANTEN Telescope $^{12}$CO(J=1-0) Data}

$^{12}$CO(J=1-0) observations were made with the NANTEN 4m telescope, located at the time in Las Campanas Observatory, Chile. The telescope front end was a 4 K cryogenically cooled SIS mixer receiver, which provided a typical system noise temperature of $\sim$250 K at 115 GHz (in a single side band), including the atmosphere towards the zenith \citep{ogawa90}. The back end was a 2048 channel acousto-optical spectrograph with a total bandwidth of 40 MHz and an effective spectral resolution of 40 kHz, corresponding to a velocity resolution of 0.1 km s$^{-1}$. \par

We utilise F99's dataset, supplementing it with archival data from \citet{matsunagaphd} and the NANTEN Galactic Plane Survey (GPS, unpublished). The inclusion of these latter observations ensures that our present dataset covers the full extent of the supershell. The $^{12}$CO(J=1-0) line occurs at 115.3 GHz, for which the telescope half power beam width is $\sim$2.6 arcmin. All observations were made by position switching, and pointing centres were arranged in a square grid with spacings of 2 arcmin or 4 arcmin between adjacent positions. The coverage and grid spacings are shown in Fig.~\ref{obsregions}, in which the regions labelled 1, 2 and 3 refer to the observations of F99, \citet{matsunagaphd}, and the NANTEN GPS, respectively. Region 1 was covered first at a grid spacing of 8 arcmin, with followup observations at 2 arcmin spacing made towards areas at which emission was significantly detected. The remaining area was later covered at 4 arcmin spacing. Region 2 was initially covered at 4 arcmin spacing, with high sensitivity 2 arcmin observations later made of the region covering the giant molecular cloud (GMC) G288.5+1.5 \citep{matsunagaphd}. Region 3 was covered at 4 arcmin only. The rms noise per channel is $\sim0.5$ K, $\sim0.3$ K and $\sim0.8$ K for regions 1, 2 and 3 respectively. 

\section{Results}

\subsection{Overview of the Neutral Gas Distribution}
\label{overview}

Fig.~\ref{COHIii} shows the Parkes SGPS H{\sc i} data overlaid with NANTEN $^{12}$CO(J=1-0) contours, integrated over the full velocity range of the Carina Flare supershell. The H{\sc i} shows a roughly elliptical cavity, bordered at its lower extremes by the bright emission of the Galactic plane, and by partial, curved walls to the left and right. Clumpy, loop-like substructure can be seen projected inside, and emission from the upper-left regions of the image is noticeably weak. The CO clouds are sparsely scattered across the field and generally show a good large-scale correlation with the integrated H{\sc i} intensity. The majority of the emission is concentrated in the GMC at $(l,b)\approx(289^\circ,1.5^\circ)$, which coincides with part of the cavity's bottom rim. With this exception, it is notable that much of the remaining molecular gas is coincident with structures projected inside the cavity, rather than its outermost walls. Several prominent molecular clouds are also seen at high latitudes, in the region in which H{\sc i} integrated intensity is weak.\par 

\subsection{Large-Scale Morphology and Dynamics of the Atomic ISM}
\label{himorph}

\begin{figure*}
\includegraphics{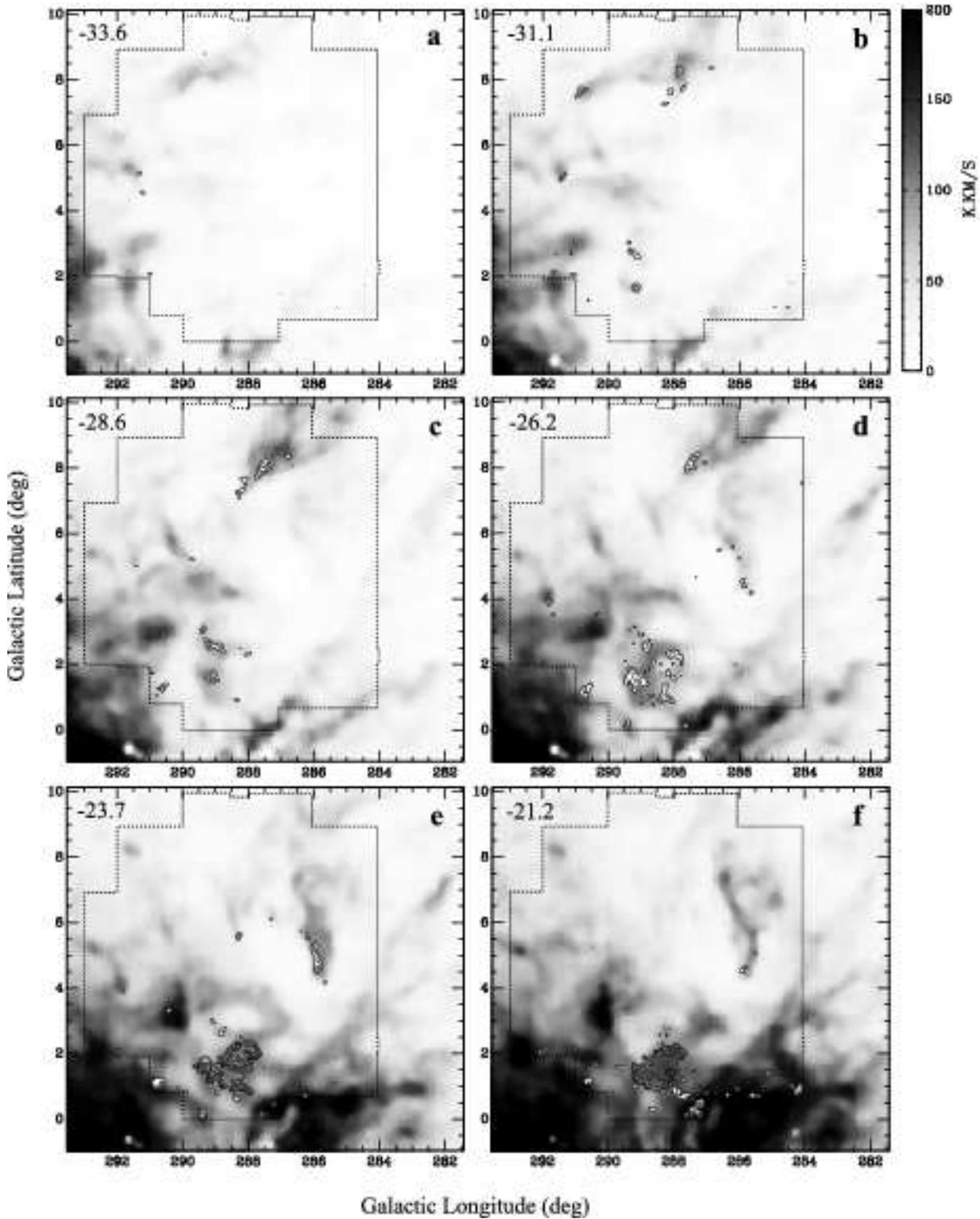}
\caption{Velocity channel maps of the Carina Flare supershell, showing Parkes SGPS H{\sc i} (grayscale) overlaid with NANTEN $^{12}$CO(J=1-0) (contours). Each panel displays an integration over three H{\sc i} velocity channels, corresponding to an interval of 2.46 km s$^{-1}$. The central velocity of each map is displayed in units of km s$^{-1}$ at the upper left. The NANTEN observed regions are marked with dotted lines. The lowest CO contour level is 1.5 K km s$^{-1}$ and the contour interval is 5 K km s$^{-1}$. The contours of the GMC in panels e and f are left unfilled to avoid obscuring the underlying grayscale.} 
\label{chsa}
\end{figure*}

\addtocounter{figure}{-1}

\begin{figure*} 
\includegraphics{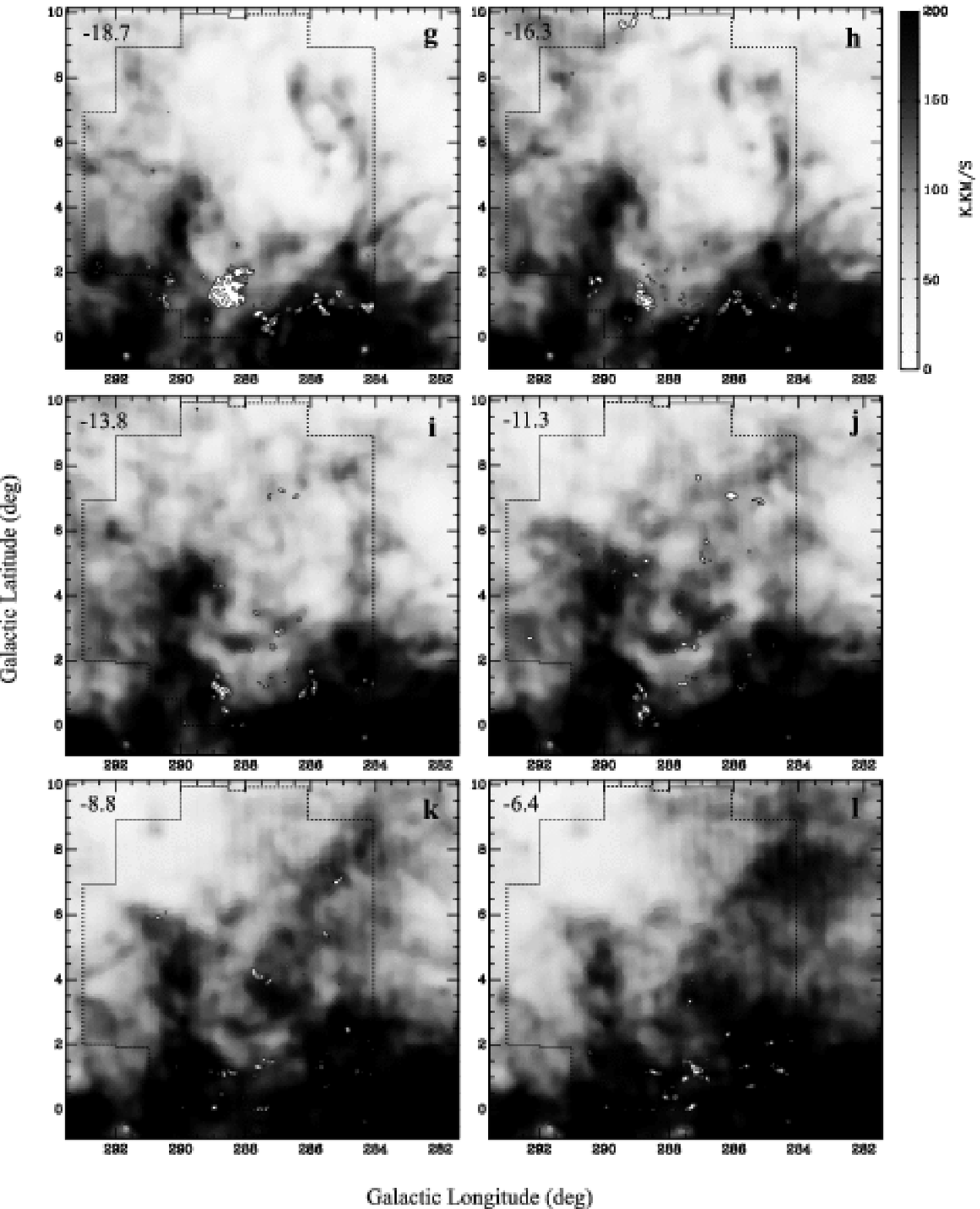}
\caption{ -- cont.}
\end{figure*}

We turn to velocity channel maps to examine the morphology, sub-structure and dynamics of the supershell in detail. Fig.~\ref{chsa} presents the two datasets, with each image displaying an integration over three H{\sc i} velocity channels (corresponding to 2.46 km s$^{-1}$). As will be discussed presently (see \S\ref{contam}), at these longitudes the contribution from physically unrelated Galactic emission becomes increasingly important as velocity approaches values close to zero. This is reflected in the maps, which show an unmistakable brightening of the entire field as $v_{lsr}$ becomes less negative.\par 

\begin{figure}
\includegraphics{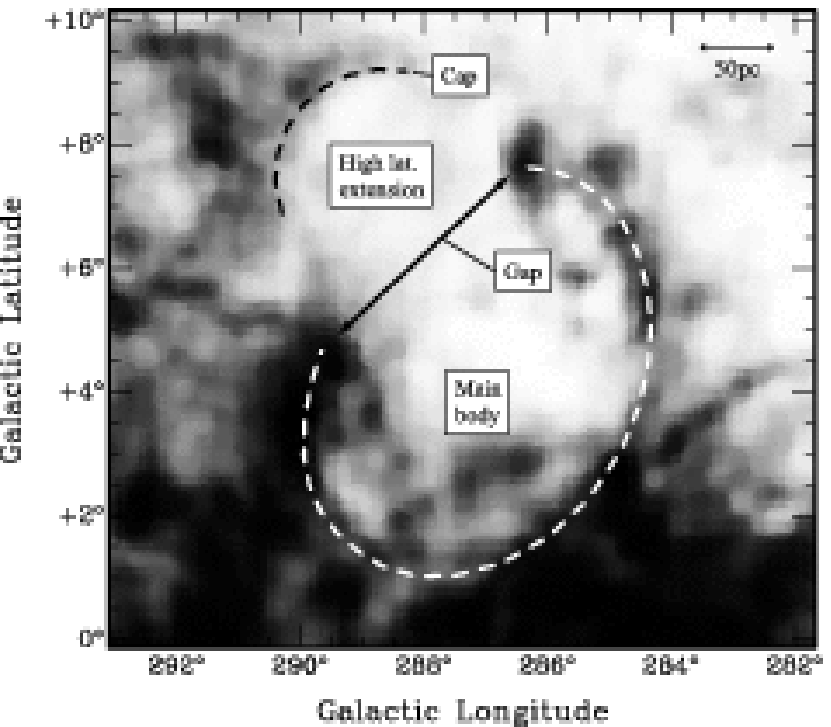}
\caption{Single H{\sc i} velocity channel map centred on -17.9 km s$^{-1}$, marked with main morphological components of the Carina Flare supershell. The emission has been artificially enhanced by a factor that scales linearly with latitude, such that $T'=(1+b/20)\ T$, where $b$ is in degrees. This brings out detail in high latitude regions.} 
\label{features}
\end{figure}

The void is seen at its widest extent in the maps centred on -18.7 and -16.3 km s$^{-1}$ (panels \textit{g} and \textit{h}). Fig.~\ref{features} shows a single H{\sc i} velocity channel at -17.9 km s$^{-1}$ on which the main morphological components of the Carina Flare supershell have been marked. The supershell may be divided into two distinct sections - a main body and a high latitude extension. The main body is well described by a partial ellipse of eccentricity $\sim0.8$, centred on $(l,b)\approx(287^\circ,4^\circ)$, measuring $5\times8^\circ$ ($\sim230\times360$ pc at $D=2.6$ kpc), with its major axis is inclined to the right at an angle of $\sim35^\circ$ to the vertical. This elliptical cavity extends down to $b\approx0.5^\circ$, and its bottom wall, below around $b\approx3^\circ$, is defined where it borders the bright emission of the Galactic plane. The cavity thus extends $\approx1.5^\circ$ ($\sim70$ pc) lower than the data coverage in F99.\par

The right-hand side of the main body is delineated by a curved, filamentary wall extending from $(l,b)\approx(284^\circ,3^\circ)$ to $(286^\circ,8^\circ)$, corresponding to an extension of $z\sim360$ pc above the Galactic plane. This wall appears relatively thin, with a characteristic projected width of $<1^\circ$ ($\sim45$ pc) at the observed resolution of $\sim16$ arcmin. A thicker, brighter spur marks the left-hand edge, protruding from the Galactic plane at $l\approx290^\circ$ and ending abruptly at $b\approx5^\circ$.\par 

The walls of the main body form a continuous ring, save for a prominent gap in its upper left edge. This gap runs from $(l,b)\approx(289.5^\circ,5.0^\circ)$ to $(286.5^\circ,8.0^\circ)$, and is almost 200 pc wide, with its centre located $z\sim280$ pc. The high latitude extension protrudes upwards and left of this gap. Here the low brightness cavity extends almost $3^\circ$ ($\sim130$ pc), and we detect a faint, curved `cap' at $(l,b)\approx(289.5^\circ,9.0^\circ)$, corresponding to an altitude of $z\sim400$ pc above the Galactic Plane. The exact extent of this cap is not clear, and it appears to be somewhat blended with emission from the surrounding region ($l\sim291.5^\circ$, $b\sim9^\circ$). Nevertheless, it shows a coherent, rounded inner edge, a sharp contrast with the void, and forms a morphologically consistent part of the shell, which leads us to believe the detection is robust. We return to the interpretation of this high latitude structure in \S\ref{blowout}, in which we suggest that Carina Flare supershell has blown out of the Plane and is in the process of forming a Galactic chimney.\par 

F99 estimated the systemic velocity of the supershell as $\sim-20$ km s$^{-1}$, based on the velocity distribution of the molecular component alone. Combined with H{\sc i}, it becomes clear the the shell's main body is at its widest and most clearly defined several km s$^{-1}$ positive of this value. We therefore refine the estimate slightly to $\sim-17$ km s$^{-1}$. This figure is also in good agreement with the values suggested by examination of the shell's structure in the $l$-$v_{lsr}$ plane (see \S\ref{fitting})

At velocity channels to either side of $v_{lsr}\sim-17$ km s$^{-1}$ emission appears interior to the void.
Approaching limb emission is observed in panels \textit{b} to \textit{f} ($-32.3 < v_{lsr} < -20.0$ km s$^{-1}$), and is generally quite sparse, showing distinct substructure. In contrast, emission at receding limb velocities ($-15.0 < v_{lsr} < -5.2$ km s$^{-1}$) appears clumpier, brighter and more heavily filled, reflecting the large contribution from physically distant gas. 

\subsection{Contamination from Unrelated Emission}
\label{contam}

\begin{figure} 
\centering
\includegraphics{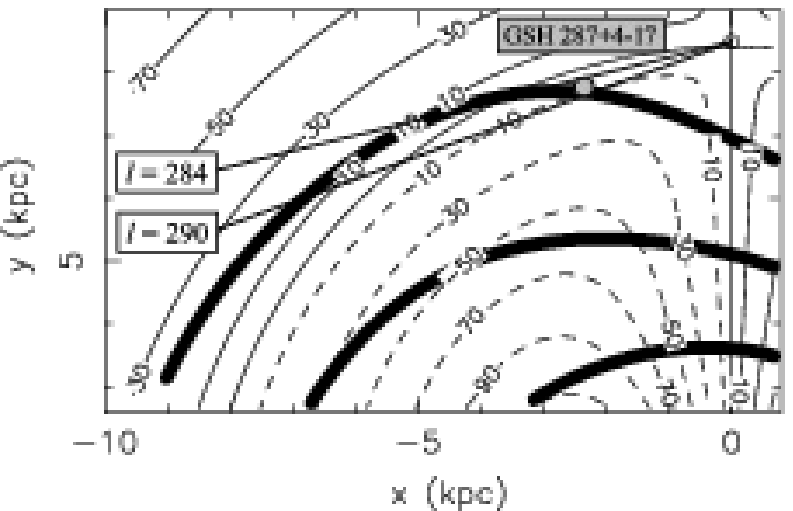}
\caption{Adapted from figure 4 of \protect\citet{mcclg01}. Velocity contours calculated from the rotation curve of Fich, Blitz and Stark, plotted together with the spiral arm pattern from \protect\citet{taylor93}. Unbroken contours represent positive velocities and dashed contours represent negative velocities. Both are labelled with $v_{lsr}$ in units of km s$^{-1}$. The Solar system is indicated by the open circle to the upper right, and the Carina Flare supershell is indicated by the filled grey circle. The longitude extremes of the shell are marked.}
\label{naomifig}
\end{figure}

\begin{figure} 
\centering
\includegraphics{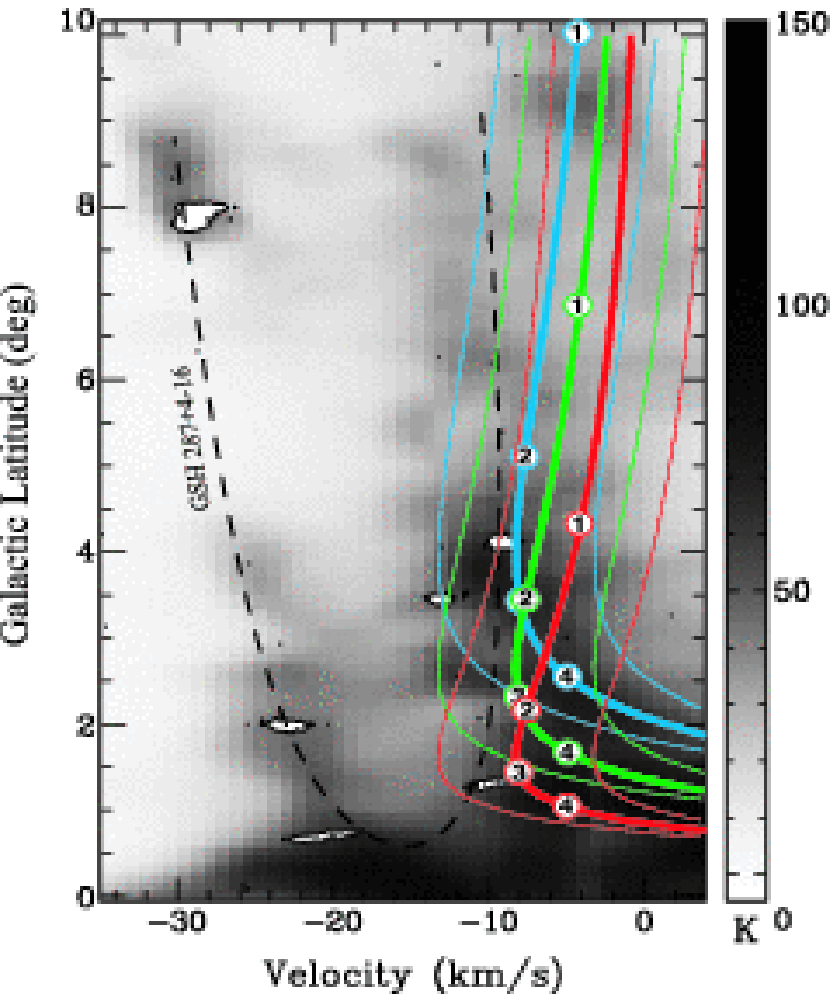}
\caption{$b$-$v_{lsr}$ slice taken at $l=287.63^\circ$. The grayscale image is H{\sc i} and filled contours are $^{12}$CO(J=1-0) at a level of 1.5K. The dashed line marks the approximate location of the Carina Flare supershell. Thick red, green and blue line are the solutions to Fich et al's rotation curve at $z=75$, 120 and 180 pc respectively. Numbers in filled circles are distances in kpc. Thin lines are plotted at velocities of 5 km s$^{-1}$ to either side of the main solutions, and are intended to illustrate a reasonable ISM velocity dispersion.}
\label{bvslice}
\end{figure}

The Carina Flare supershell lies close to the tangent of the Carina Arm, at $l\approx280^\circ$. The velocity structure in this region is such that for certain values of $v_{lsr}$ the signal from the supershell is heavily confused with emission from physically distant gas. Fig.~\ref{naomifig} shows the shell's position marked on an isovelocity contour plot calculated from the rotation curve of \citet{fich89}, overlaid with the spiral arm pattern of \citet{taylor93}. At the longitude of the supershell the line of sight is dominated by emission in the range $-10 \la v_{lsr} \la 10$ km s$^{-1}$; mainly originating from a column extending several kiloparsecs through the arm, but also including contributions from the interarm region and immediate Solar vicinity. This fact is borne out by the observational data, which shows multiple bright and complex components in this range. \par

The localised region of the near side of the Carina Arm exhibits well-known departures from circular rotation \citep{alvarez90, georg00, russeil03}, which result in much of the gas possessing values of $v_{lsr}$ more negative than permitted by standard rotation curves. The supershell's velocity range reflects this; covering $-36 \la v_{lsr} \la -3$ km s$^{-1}$ at longitudes for which $v_{lsr}\la-12$ km s$^{-1}$ is generally `forbidden'. This negative offset is fortuitous, and means that the forward expansion of the approaching limb carries it to values of $v_{lsr}$ that are comfortably removed (by $>10$ km s$^{-1}$) from most of the Galactic signal. The emission can therefore be placed in the immediate vicinity of the shell with a high degree of certainty.\par

The same is not true for the receding limb. Fig.~\ref{bvslice} shows a velocity-latitude plot of the H{\sc i} and CO datasets, taken at the approximate centre of the supershell. Solutions to Fich et al's rotation curve are overlaid on the image, computed for constant heights above the galactic plane of $z=75$, 120 and 180 pc, with distances marked. The 75 pc and 180 pc lines are chosen to represent the density scale heights of the molecular and atomic discs respectively \citep[and references within]{dickey90, malhotra94a}, where we compute an effective scale height given by $H=\int_0^\infty{\frac{n(z)}{n_0}}\ dz$ \cite[e.g.][]{tomisaka98}.\par

The receding limb overlaps with the rotation curve solutions, which exist for velocities of $v_{lsr}\ga-8$ km s$^{-1}$ at the longitude of the plot. The contribution from the Galactic disc will drop off with latitude, as the line of sight passes through the lower column densities of the interarm ($D\la2$ kpc) and higher $z$ regions. However, for H{\sc i}, contamination may be present even at the highest latitudes of the shell. The molecular component, on the other hand, has a three-fold advantage: a smaller scale height, smaller internal velocity dispersions, and a distribution which is highly concentrated in the spiral arms. For a gaussian distribution with $H\sim75$ pc, $\sim95$ per cent of the mass is found below the $2\sigma$ point at $z\approx120$ pc. Below this height, emission from the spiral arm is restricted to $b\la4^{\circ}$. The arm--interarm density contrast for molecular gas in the Carina Arm has been estimated at 13:1 \citep{grab87}, suggesting that the contribution from the interarm regions should be minimal. In addition, the behaviour of the solutions with $z$ is such that brightest interarm emission, from close to the midplane, tends to be displaced away from the shell wall in velocity. This combination of factors suggests that the CO dataset may be largely contamination free above $b\approx4^{\circ}$, and is more reliable than H{\sc i} as a tracer of the shell's receding limb.\par 

Quantitatively assessing the H{\sc i} contribution from physically distant emission is non-trivial. The receding limb signal is superposed on a bright, complex and non-systematically varying background, and is not readily separable. However, it is illustrative to compare the velocity integrated intensities over the velocity ranges of the approaching and receding limbs. The ratio is typically $\sim1:4$, suggesting that the signal from the supershell may only contribute a few 10 per cent of the total emission observed at the velocities of its receding wall.  

\subsection{Comparison of Parkes SGPS H{\sc i} and NANTEN $^{12}$CO}
\label{hico}

The H{\sc i} and CO clouds generally appear to form well-associated, co-moving components of the supershell. The Parkes SGPS data also newly reveal rich substructure in the atomic ISM, and some interesting morphological relationships with the dense molecular phase. The clearest examples of H{\sc i}-CO association are seen on the contamination-free approaching limb (panels \textit{a} to \textit{f} in Fig.~\ref{chsa}), and we briefly describe several of the most notable features here: 

\textit{i)} The lower left edge of the main body is delineated predominantly by molecular gas, in the form of the bright ridge of the GMC G288.5+1.5. This ridge is seen between $(l,b)\approx(288.0^\circ,0.5^\circ)$ and $(289.5^\circ,2.0^\circ)$, and is most prominent in panels \textit{e} and \textit{f} ($-24.7 < v_{lsr} < - 20.0$). We return to this feature briefly in \S\ref{mcformn}.

\textit{ii)} A pair of bright H{\sc i} filaments are seen in panels \textit{d} to \textit{f}. These run from $(286.5^\circ,7.5^\circ)$ and $(286.5^\circ,6.0^\circ)$ respectively, and converge to a common point at $(285.8^\circ,4.5^\circ)$. They curve parallel to the right-hand wall, projected inside the main body, and CO emission delineates the inner of the two, extending almost 2$^\circ$ ($\sim$90 pc) in length. A bright star-forming clump is located at the point where the two converge \citep{dawson07}. 


\textit{iii)} Highly blueshifted emission in the top left of panels \textit{a} and \textit{b} is well interpreted as fast-approaching components of the high latitude extension of the supershell. The H{\sc i} arc between $(l,b)\approx(291.0^\circ,7.5^\circ)$ and $(288.0^\circ,8.0^\circ)$, and its associated molecular cloud, fall in a similar position to the cap observed in central channels. 

\textit{iv)} The most prominent high latitude structure is the bright H{\sc i}-CO complex at $(l,b)\approx(287^\circ,9^\circ)$, visible in panels \textit{b} to \textit{d}. This striking, morphologically unusual structure begins with a sharp pointed tip at $(288^\circ,7.5^\circ)$, towards which the molecular component is concentrated, but flares out to the right, extending over 4$^\circ$ ($\sim200$ pc) in length. Its relation to the large-scale morphology of the supershell is unclear, but the presence of so much molecular gas so far above the Galactic plane ($z\sim350$ pc) is notable.\par 

These features showcase a variety of morphological relationships between the atomic and molecular ISM, and much of the observed structure strongly suggests the formative influence of the supershell. This marks it as an extremely promising target for in-depth investigations into the impact of supershells on the evolution of the ISM, and in particular, their proposed role as molecular cloud producers. We return to this topic briefly in \S\ref{mcformn}, but plan to address it in detail in forthcoming papers, which will compare the two phases at matched resolutions of $\sim2$-3$'$.\par 


In the receding limb, the high level of H{\sc i} contamination means that clear associations between individual H{\sc i} structures and molecular clouds are less readily observed. Nevertheless, the two phases are still broadly correlated. In panels \textit{i} to \textit{l} a wide scattering of small molecular clouds can be seen in the region of the main body, associated with areas of bright H{\sc i} emission. The peak intensity of this molecular cloud component occurs at $\sim-10$ km s$^{-1}$, vanishing almost entirely by $\sim-6$ km s$^{-1}$, whereas H{\sc i} emission peaks typically occur at more positive values. Much of the atomic emission must be interpreted as arising from gas physically distant from the supershell (see \S\ref{contam}), and as the more reliable tracer of genuinely associated emission, the CO velocity peak is likely to be the more useful figure here.\par

Finally, it is also instructive to examine H{\sc i} linewidths for the information they may provide on the location of cool atomic gas within the shell walls. H{\sc i} line profiles are most easily observable in structures on the approaching limb, and reach (FWHM) values of as low as 4-6 km s$^{-1}$. The narrowest profiles tend to occur close to or coincident with CO emission, but not exclusively so, with values of 5 km s$^{-1}$ found over a degree from molecular clouds. Under the assumption of pure thermal broadening, these values correspond to kinetic temperatures of $350\sim800$ K. In reality, turbulence is likely to contribute significantly to the observed velocity dispersion, and such low linewidths indicate relatively cool H{\sc i} within the shell wall. Forthcoming high resolution studies will offer a means of localising narrow-line features in the shell walls, facilitating more meaningful comparisons with the CO distribution.

\begin{figure*} 
\centering
\includegraphics{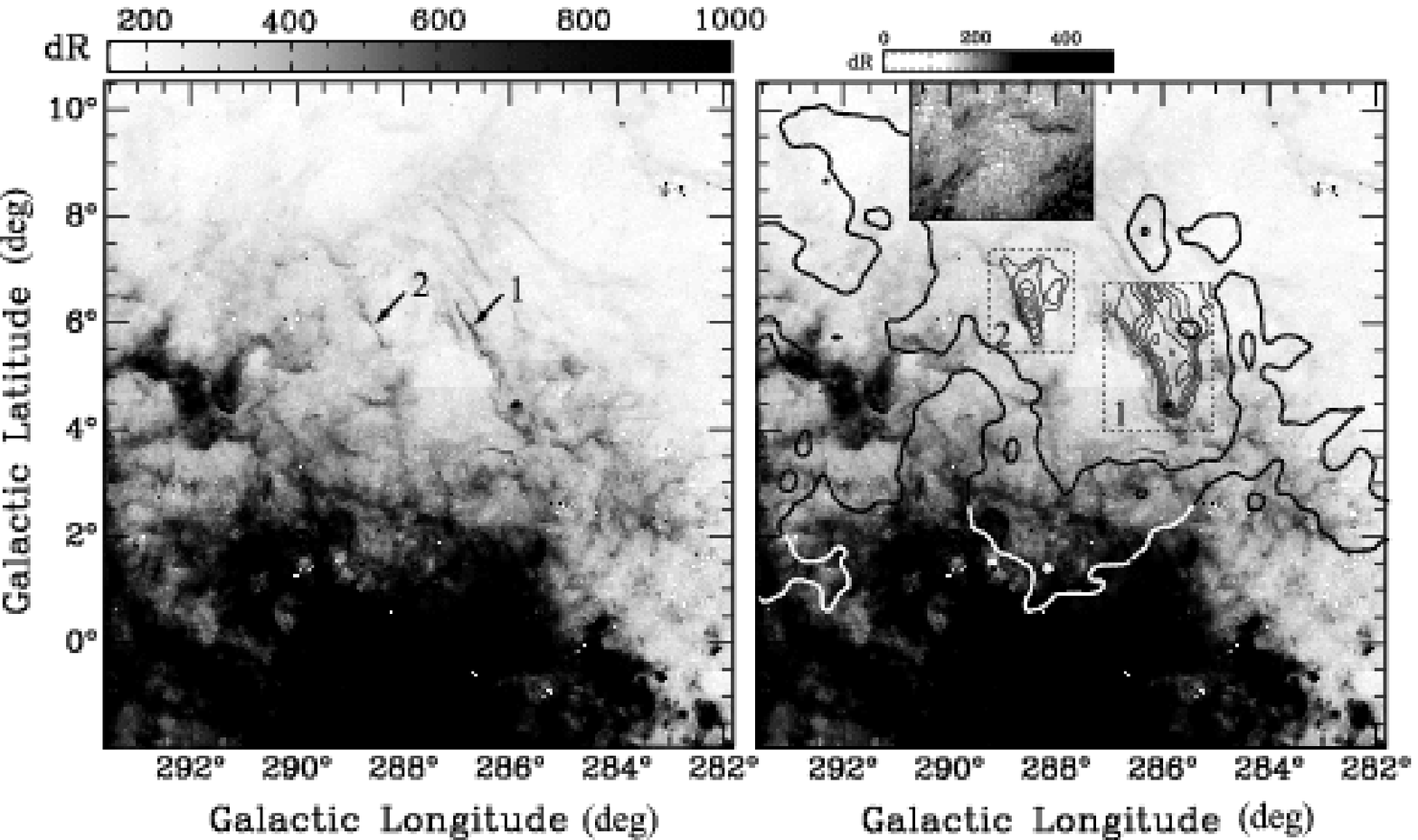}   
\caption{Grayscale: SHASSA H$\alpha$ image of the Carina Flare region. The scale is chosen to bring out faint details in the high latitude portions of the image. Left: Filaments that show clear associations with H{\sc i} are labelled 1 and 2. Right: Thick contours show H{\sc i} in a single velocity channel centred on -17.1 km s$^{-1}$, with contour levels of 20K and 50K. These are coloured black or white as appropriate, to stand out against the background. The sub-region bounded by a solid box is contrast-enhanced to bring out the faint H$\alpha$ cap feature. Sub regions bounded by dotted lines, labelled 1 and 2, show H{\sc i} features associated with the H$\alpha$ filaments. Velocity integration ranges and contour levels are (1) $-28.6 < v_{lsr} < -17.1$ km s$^{-1}$, $170+27$K km s$^{-1}$ and (2) $-27.0 < v_{lsr} < -22.0$ k ms$^{-1}$, $100+13.5$K km s$^{-1}$ respectively.}
\label{shassaall}
\end{figure*}

\subsection{Mass of the swept-up ISM}
\label{mass}

The mass swept up by a shell is an important parameter in determining its age and energetics, as well as the characteristics of the medium into which it expanded. The Parkes H{\sc i} data allow for significantly improved mass estimates for both the atomic and molecular components; the latter through improvements in the accurate identification of associated molecular matter. A distance of $2.6\pm0.4$ kpc is assumed throughout, which will be fully justified below (see \S\ref{location}).

\subsubsection{Atomic Hydrogen}

The atomic mass swept up in the supershell may be estimated from the H{\sc i} brightness-column density relation, $N_\mathrm{HI}=1.8\times10^{18}\int{T_{b }\ dv}$ \citep[e.g.][]{dickey90}, which holds in the optically thin limit. At the velocities of the receding limb the signal from the supershell is heavily confused with the contribution from non-local Galactic emission, from which it cannot be reliably disentangled. We therefore utilise the approaching limb emission alone, which is relatively free from contamination and often shows well-defined gaussian line profiles. We consider only components away from the shell edges so as to minimize curvature effects, and separate them from the background by decomposing the line profiles into multiple gaussians. We then obtain estimates for the average column density through the front wall in 20$'$ latitude intervals and convert these to mass contributions by assuming that the radius of the shell in the third spatial dimension is equal to its width in longitude, which we measure in the $l$-$b$ and $l$-$v_{lsr}$ planes. The atomic mass is thus estimated to be $M_\mathrm{HI}\approx7\pm3\times10^5~\mathrm{M}_{\odot}$, where the quoted uncertainty accounts for measurement errors on the longitude widths and the uncertainty in the distance.\par 

\subsubsection{Molecular Hydrogen}

The mass of associated molecular hydrogen is estimated from the widely used empirical relation between $^{12}$CO(J=1-0) velocity-integrated intensity, $W_{\mathrm{CO}}$, and molecular hydrogen column density, $N_{\mathrm{H}_2}=XW_{\mathrm{CO}}$. In keeping with F99, we adopt a value of $X = 1.6\times10^{20}$ K$^{-1}$ km s$^{-1}$ cm$^{-2}$ \citep{hunter97}. \par

F99 chose a conservative estimate of the related molecular gas, excluding most of the GMC from their data coverage and analysis. Now, the combined Parkes SGPS and NANTEN data confirm its association with the supershell, enabling us to improve upon their mass estimate. In total, $\sim85$ per cent of the GMC emission appears to belong to the Carina Flare, with the remainder better explained as related to the Carina OB2 supershell. These components are generally separable in the $l$-$v_{lsr}$ plane (see e.g. panel \textit{a} of Fig.~\ref{fits}), where they are seen to follow the edges of one or other of the H{\sc i} cavities created by each shell. The remainder of the molecular clouds are well described as discrete islands in the datacube that are small compared to the dimensions of the supershell. We include a cloud in the mass estimation if the $l$, $v_{lsr}$ coordinates of its peak at any given latitude place it within the range of the supershell, as estimated from the fits described in appendix \ref{fitting}. Although the limitations of the fits will be discussed, they are sufficiently accurate for a first approximation. In practice, $\ga80$ per cent of the total molecular emission is contained in the GMC and several robustly associated objects on the approaching limb, and the exact inclusion criteria have a minimal effect on the final mass.  We thus obtain $M_{\mathrm{H}_2}\approx2.0\pm0.6\times10^5~\mathrm{M}_{\odot}$, where the uncertainty arises from the distance estimate. This improved figure is almost 3 times larger than that of F99. \par

\subsubsection{Number density of the Ambient Medium}

By assuming a simple 3D shape for the shell we may estimate the average number density of the medium into which it expanded. The swept-up mass is contained in two components -- the $\sim230\times360$ pc main body, and the smaller high-latitude extension. Assuming the depth of the shell is close to its width, and crudely representing the high-latitude component as half of an ellipsoid with dimensions of $\sim180\times180\times360$ pc, we find that an average H{\sc i} number density of $\sim3$cm$^{-3}$ would be required for the supershell to have swept up all of its estimated mass from an ambient atomic ISM. 

\subsection{Comparison with SHASSA H$\alpha$ emission}
\label{ionized}

Comparisons with tracers of H{\sc ii} can provide valuable information on the ionized content of a supershell. No such comparison has yet been performed for the Carina Flare supershell. We utilise data from the Southern H$\alpha$ Sky Survey Atlas \citep[SHASSA, ][]{gaustad01}. The images are continuum-subtracted, with a pixel size of 0.8 arcmin, and are smoothed to a resolution of 4.0 arcmin, resulting in a sensitivity of $\sim0.5$R.\par 

Fig.~\ref{shassaall} shows a SHASSA H$\alpha$ image of the supershell region, with the grayscale chosen to bring out high latitude structure. Relevant H{\sc i} features are also marked. The field is dominated by the Carina Nebula at $l=287.6^\circ$ $b=-0.9^\circ$, and bright emission from the Galactic Plane drowns out much of the region at $b\la2^\circ$. The lower edge of the supershell shows good positional agreement with the edge of this bright emission, and the dense ridge of the GMC is seen in absorption, indicating that at least some of the ionized gas lies behind it.\par

The field shows a wealth of filamentary structure. Below $b\approx4^\circ$ the emission is highly confused, and in the absence of velocity information for the H$\alpha$ line, we are unable to relate it to the Carina Flare. However, between $4^{\circ} < b < 8^\circ$, where confusion is lower, delicate, filamentary strands of ionized gas are seen that show an intriguing correspondence with the supershell. 
These streamer-like filaments run nearly straight and remarkably parallel. They are projected inside the void, mainly falling in the region of the gap in the main body, and are oriented such that they appear to be pointing directly out through the gap and into the shell's high latitude extension. The two brightest, labelled `1' and `2' in the figure, align neatly with the edges of well-identified HI structures on the approaching limb, confirming a genuine association with the supershell. This makes the Carina Flare supershell one of only a handful of galactic shells for which an ionized component has been positively identified. Other fainter filaments also show suggestive associations, and there is evidence for a faint ionized counterpart to the H{\sc i} cap. This is shown in a contrast-enhanced portion of the image, and appears to follow the line of curvature of the the atomic gas feature. The ionized gas in the streamer-like filaments is generally seen along the lower left rims of HI clouds and the features as a whole are concentrated towards the top left of the shell. Their projected lengths and widths are $50\sim100$ pc and $2\sim5$ pc respectively (the latter at the resolution limit of the smoothed images).\par

The origins and physics of this ionized component will be the subject of future papers, as will a comparison with the stellar population in the region. However, even considering morphology alone, the ionized gas features may provide clues to the evolutionary state of the supershell. This is discussed further in \S\ref{blowout}.

\section{Discussion}

\subsection{The Carina Flare supershell in the Galactic Environment 1: Location}
\label{location}

F99 assign a distance of $2.6\pm0.4$ kpc for the Carina Flare. This figure was based mainly on known distances to major molecular clouds and OB stars in the shell longitude range, and was also supported by a kinematic distance estimate. This kinematic estimate was derived for $v_{lsr}=-20$ km s$^{-1}$, assuming old IAU standard values for the galactocentric radius and orbital velocity of the solar system ($R_0=10$ kpc, $\Theta_0=250$ km s$^{-1}$). However, under the current recommended values ($R_0=8.5$ kpc, $\Theta_0=220$ km s$^{-1}$), velocities of $v_{lsr} \la-12$ km s$^{-1}$ fall outside the solution range of circular rotation models at this longitude \citep{fich89, brand93, clemens85}, and the result is not reproducible. This urges a careful re-examination of the distance to the supershell. Fortunately, the Parkes H{\sc i} dataset enables improved discussion of the Carina Flare's place in the local Galactic environment.\par

As discussed above, the near side of the Carina Arm, covering the region $280 < l < 310^{\circ}$, exhibits departures from circular rotation that result in a characteristic negative offset in observed $v_{lsr}$ values (see \S\ref{contam}). The arm feature has been well traced in both H{\sc i} and CO \citep{grab87}, and is clearly visible in our SGPS dataset as a bright ridge in the $l$-$v_{lsr}$ plane. The supershell's velocity places it within this bright ridge, where it is robustly identified as a void of reduced emission, strongly implying a genuine association with the arm.\par 

\citet{georg00} use data from the literature to determine distance moduli to the exciting stars of H{\sc ii} regions between $285 < l < 293^\circ$, and find that objects with `anomalously' negative line-of-sight velocities are confined to distances of $2.5 \la D \la 3$ kpc. This is consistent with estimates of distances to the Carina arm feature from spiral arm models \citep[e.g.][]{cordes02,russeil03}, also also with the estimated distance to the stellar density peak ($\sim2.7$ kpc) used by F99. We also note that the GMC G288.5+1.5 is seen in absorption against the bright H$\alpha$ emission that dominates the Galactic plane at $b\la2^\circ$. \citet{georg00} identify most of this emission as part of the Carina OB2 supershell \citep{rizzo98}, for which they calculate a distance estimate of 2.9 kpc. Other studies place the central cluster at $D=3.1\pm0.3$ kpc \citep{garcia94}. This appears to provide an upper limit of $D\approx3$ kpc for the Carina Flare supershell.\par

We may also consider whether the supershell's morphology and velocity structure offer any clues to its placement in the local Galactic environment. The main body of the Carina Flare forms an almost perfect ellipse, but is inclined at an angle of $\sim35^\circ$ to the vertical, indicating that at some stage expansion proceeded preferentially in a direction that included a horizontal component -- rightwards on our images. As we will discuss below (\S\ref{blowout}), the vertical disc density gradient plays a crucial role in a supershell's evolution. However, the effect of the arm--interarm density contrast may also be of comparable magnitude, encouraging enhanced expansion into the interarm regions \citep{mcclg02}. We speculate that the enhanced horizontal component of the Carina Flare supershell's expansion may be the result of such a density gradient. If so, this places the shell on the near edge of the Carina Arm. (That is, the clockwise side as viewed from the Galactic North.) In such a location, the central density peak of the Arm occurs to the left, as viewed in the orientation of our images, and the negative density gradient includes a large component to the right. This may be readily confirmed on Fig.~\ref{naomifig}.

Exploring this idea further, we consider that supershells formed by OB associations are likely to form initially in the spiral arms where such young objects are found. Depending on their location in the Galaxy, they may then migrate out into the interarm regions on time scales comparable to their lifetimes. The length of time a shell remains in its parent arm depends on the difference between the rotation speed of the gas and the speed of the spiral pattern, $\Omega_p$, where $\Omega_p$ is a function of Galactic radius. The spiral pattern speed is not well determined, but recent studies suggest that $\Omega_p\sim20\pm5$ km s$^{-1}$ kpc$^{-1}$ \citep[see][and references within]{leitch98}. The Carina Flare supershell is located in the inner galaxy, at $R\sim8$ kpc, where the spiral pattern speed is therefore estimated as $\sim160$ km s$^{-1}$. For a Galactic rotation speed of 220 km s$^{-1}$, the resultant slow migration of the shell through the spiral arm occurs in a direction towards the front edge of the arm, at a relative speed of $\sim60~{\rm pc~Myr^{-1}}$. Under these assumptions, the shell progresses several hundred parsecs through a $\sim2$ kiloparsec arm within $\sim10^7$ years (see \S\ref{eanda}), but it need not have emerged completely, and will not have had time to encounter a second spiral shock. The hypothesised location on the near edge of the arm therefore appears to be quite reasonable.\par 

It should be noted that we do not consider this scenario to be at odds with the interpretation that Galactic Plane blowout has occurred to the upper left, as will be discussed below. The direction in which hot gas first begins to flow out of the shell interior will be determined by the exact location at which the shell first begins to fragment, which may depend strongly on the local configuration of the ISM both in the shell and the high latitude regions. \par

In light of all the factors presented above, we have opted to retain a value of $D=2.6\pm0.4$ kpc throughout this paper. 2.6 kpc appears to be a reasonable estimate of the distance to the Carina Arm feature, with which the shell is confidently associated. The 0.4 kpc uncertainty is probably a fair reflection of the tentative upper and lower limits placed on the Carina Flare supershell's distance by its relationship to the Carina OB2 supershell and its proposed location on the near edge of the arm, respectively.

\subsection{The Carina Flare Supershell in the Galactic Environment 2: Galactic Plane Blowout \& Chimney Formation}
\label{blowout}

Environment is a major governing factor in a supershell's evolution. For disc galaxies such as the Milky Way, the disc density gradient becomes important when a supershell grows to exceed the gas scale height. Accelerated expansion along the gradient away from the Plane leads to the growth of Rayleigh--Taylor instabilities, eventually culminating in the break-up of the shell and the release of its interior gas \citep{avillez01, maclow89, ttagle90}. This hot, overpressurized material streams between the shell fragments and flows upwards, forming a galactic `chimney', through which energy and enriched material are supplied to the Halo \citep{norman89}. Such supershells are then often said to have `blown out' of the Plane. Hydrodynamical simulations find that Galactic Plane blowout typically occurs at a height of between $1\la H\la3$ for a range of initial density distributions \citep[e.g.][]{avillez01, igumen90, maclow89}, where $H$ is the effective scale height, $H=\int_0^\infty{\frac{n(z)}{n_0}}\ dz$. These predictions have been supported by recent observations of a number of known Galactic chimneys \citep{mcclg03, mcclg06, callaway00}.\par

The Carina Flare, with its broken main body, `puffed out' high-latitude extension and faint cap, shows convincing morphological evidence for Galactic Plane blowout. The gap in the main body is located at $z\sim280$ pc, which corresponds to $\sim1.6H$ in the H{\sc i} vertical density distribution of \citet{dickey90}. This figure is comfortably consistent with expected fragmentation heights, and we suggest a scenario in which the supershell's hot interior gas has broken out of the main body and is flowing out through the gap to the upper left. The cap is interpreted as a structure swept up from the material of the upper disc; a phenomenon that is also observed in model chimneys \citep{igumen90, maclow89, ttagle90}. We note that similar structures have also been studied observationally in the exceptionally large Galactic chimney GSH 242-03+37, which shows several high-latitude ($z\sim1600$ pc), low-brightness arcs, capping the high-$z$ extensions of chimney outflows \citep{mcclg06}.\par

The blowout is apparently one-sided, with no obvious indication of counterpart bubble expansion on the bottom side of the disk, implying that the parent stellar cluster was displaced from the Galactic Plane. Numerical simulations suggest that such one-sided chimneys are easily produceable given a realistic distribution of OB clusters. Models of individual supershells in idealised density distributions suggest that an initial cluster height of $z\sim100$ pc is sufficient to produce one-sided expansion \citep{maclow88, maclow89, tomisaka86}. This is further supported by \citeauthor{avillez01}'s (2001) large-scale, multi-shell simulations of the Galactic Disc, which produce one-sided chimneys using an OB cluster distribution with a scale height of 46 pc.\par

The morphology of the ionized component provides further support for the blowout scenario. The streamer-like H$\alpha$ filaments observed between $4 < b < 8^\circ$ are unusually collimated and remarkably well aligned with the proposed direction of outflow. Their projected positions place them in or around the gap; however, their association with H{\sc i} emission clearly indicates that they represent the ionized edges of parts of the shell ISM, rather than structures interior to it. Regardless, the morphological correlation is striking. We suggest that the interaction of the outflowing gas with the surrounding supershell walls provides a means of producing such elongated, collimated structure. Such associations are not unprecedented: similar structures have been observed in the small, H$\alpha$-bright LMC supershell DEM152, where they have been definitively associated with a confirmed point of blowout \citep{chen00}. Here too, the authors find rich, filamentary H$\alpha$ structure, projected through the break in the shell walls, and forming delicate strands running parallel to the direction of outflow.\par

At present it is not clear whether the Carina Flare supershell forms a fully open channel to the Halo, or remains confined in the upper regions of the disc. The cap feature is only confidently detected over a limited spatial area and does not appear to fully enclose the high latitude extension. Of particular interest is an apparent break detected at $(l,b)\approx(287.3^\circ, 8.8^\circ)$ ($z\sim400$ pc), which is visible in Fig.~\ref{features}. However, emission appears in this location in positively adjacent velocity channels, and given the high background levels, this cannot be either confirmed or ruled out as part of the shell. Direct observations of the hot gas itself would help to clarify the extent of the outflow, providing a means of confirming the evolutionary scenario presented here. \par

Nevertheless, it should be stressed that 
ultimately, whatever the detailed evolutionary status of the supershell, several important features can be stated with confidence. Firstly, it has expanded far out of the Galactic Plane, reaching heights of $z\sim450$ pc. This brings it to the vicinity of the disc--halo interface \citep[$z\sim500$ pc,][]{lockman84}, to which it should therefore already be transporting hot, enriched gas. Secondly, it is clear that its expansion has not yet stalled. Both simulations and observations therefore suggest that it is on the road to becoming a fully fledged chimney. 

\subsection{Energy and Age of the Supershell}
\label{eanda}

With no confirmed energy source, there is a limit to the accuracy with which the age and energetics of Carina Flare supershell can be determined. Nevertheless, the Parkes SGPS H{\sc i} data puts the present study at a significant advantage over previous work. In light of our new knowledge about the supershell's morphology, mass and evolutionary status, we are able to explore its age and formation energy in more detail than has previously been possible. \par

Beginning with quantities obtained directly from the observational data, the kinetic energy of the supershell may be simply estimated from its mass and expansion velocity. Lacking detailed knowledge of its 3-dimensional velocity structure, we assume a constant expansion velocity of $v_{exp}\sim10$ km s$^{-1}$, estimated from the combined H{\sc i} and CO datacubes. The neutral gas mass is taken as $M_{\mathrm{HI+H}_2}\sim1.3\times10^6~\mathrm{M}_{\odot}$, which includes a factor of 1.35 to account for the presence of helium and heavier elements. This results in $E_K\sim1\times10^{51}$ erg. We may also define a `dynamic age', $\tau_k=R/v_{exp}$, where $R$ is the characteristic radius of the shell. We take $R\sim140$ pc, which is the radius of a circle with the same projected area as the elliptical main body, resulting in $\tau_k\sim1\times10^7$ yr.\par 

In order to explore energy input we must turn to models of supershell evolution. The evolution of a thin, adiabatic supershell in a uniform medium has been described analytically \citep{mccray87, weaver77}. In this highly idealised system, the kinetic energy of the swept-up shell is found to be 20 per cent of the total energy input. Our kinetic energy estimate therefore implies a formation energy of a few $10^{51}$ ergs; the equivalent of several supernovae. Under the formulation of \citet{mccray87}, $R=97\mathrm{~pc~} (N_{*}/n_0)^{1/5}t_7^{3/5}$ and $v_{exp}=5.7\mathrm{~km~s}^{-1}(N_{*}/n_0)^{1/5}t_7^{-2/5}$, where $t_7$ is time in units of $10^7$ years, $n_0$ is the initial number density in cm$^{-3}$, and $N_{*}$ is the number of stars in the parent cluster with masses $\ga7~\mathrm{M}_{\odot}$. Here, each massive star is assumed to inject a mechanical luminosity of $10^{51}$ erg in its supernova explosion, at a rate of $6.3\times10^{35}N_{*}$ erg s$^{-1}$, and the parameter $N_{*}$ includes both those stars that have already gone supernova and those which are yet to explode. Solving for $R\sim140$ pc, $v_{exp}\sim10$ km s$^{-1}$ and $n_0\sim3$ cm$^{-3}$ (see \S\ref{mass}), we obtain $t_7\sim0.8$ and $N_*\sim30$. This corresponds to an energy input of $\sim5\times10^{51}$ erg. We take this opportunity to correct an error in F99. Their energy estimate exceeded ours by an order of magnitude, due to a misinterpretation of the $N_{*}$ parameter.\par

In reality, supershell evolution may deviate significantly from the idealised analytical model, and the last 20 years have seen the development of a large number of numerical models, which seek to explore a more realistic range of physical processes and environments. As direct comparators to individual objects, such models tend to be of limited usefulness, since their purpose has primarily been to explore the effects of different approaches and the inclusion of different physics. Comparisons with observational results therefore tend to be hindered by large numbers of free parameters for which observational constraints are poor. In addition, the initial configuration of the ISM (which may be highly complex and irregular) is likely to strongly affect an individual supershell's evolution \citep{avillez01, breit06}, and may result in significant differences between model predictions and reality.\par  

Nevertheless, with these limitations in mind, we may still make some broad comparisons with model results. The Carina Flare's proposed status as a young Galactic chimney is of use here, enabling us to examine energy requirements for chimney formation in order to estimate a lower limit on the mechanical luminosity injected into the supershell. Mac Low \& McCray's (1988) semi-analytic treatment finds that Galactic Plane blowout occurs above luminosities of $\sim10^{37}$erg s$^{-1}$, for ambient midplane densities of $\sim1$cm$^{-1}$ and pressures of $10^4k$ dyne cm$^{-2}$ (although their choice of model atmosphere results in shells that grow unrealistically large before blowout occurs). More sophisticated hydrodynamical models suggest similar figures, with blowout observed for input luminosities greater than a few $10^{37}$ erg s$^{-1}$ \citep[e.g.][]{igumen90, maclow89, ttagle90, tomisaka92}. For comparison, this corresponds to $N_*\ga35$ under the formulation described above. Blowout time-scales in these models are highly sensitive to the input luminosity and density distribution, and range from a few $10^6$ to a couple of $10^7$ yr for a reasonable range of parameter space. Longer time-scale, higher energy solutions tend to occur in models that include magnetic fields, which inhibit vertical shell growth \citep[e.g.][]{tomisaka92, tomisaka98}. However, more elaborate MHD simulations seem to indicate that this effect would be mitigated in a realistic, tangled field \citep{avillez05}.

\subsection{Remarks on Molecular Cloud Formation}
\label{mcformn}

The origin of the molecular material in the Carina Flare will be the topic of a forthcoming paper.
Nevertheless, it is worth commenting briefly here on the feasibility of a triggered formation scenario.

A key question is whether the shell lifetimes derived above are compatible with the formation time-scales for molecular clouds. Simulations of H$_2$ formation in large-scale flows suggest that a shock velocity of as little as $\sim10$km s$^{-1}$ can produce highly molecular gas in $\la10^7$ yr for a pre-shock number density of a couple of atoms per cubic centimetre \citep{bergin04}. This time-scale is highly density-dependent, and decreases rapidly for denser initial conditions. The Galactic ISM is a highly inhomogeneous medium that exhibits density fluctuations over length scales spanning several orders of magnitude \citep[e.g.][]{dickey01, green93}, and localised regions of enhanced density would certainly have been present in the pre-shell medium. Moreover, recent work suggests that formation time-scales become significantly lower when models are expanded to include realistic, turbulent gas dynamics \citep{glover07}. These factors suggest that molecular cloud formation over the lifetime of the Carina Flare is a viable possibility.

Obtaining convincing observational proof is a much more complex problem, and is beyond the scope of this paper. However, it is instructive to examine the case of one particularly interesting object. The giant molecular cloud G288.5+1.5 (see \S\ref{hico}) shows a bright ridge that delineates the rim of the Carina Flare at exactly the point of its projected interface with the nearby Carina OB2 supershell \citep{rizzo98}. Converging supershells provide an effective means of driving strong, large-scale compressive flows that encourage rapid molecular gas formation \citep{hartmann01}, and the location of the shell's single GMC at this point is notable. The distance estimates to the two supershells place them promisingly close (see \S\ref{location}), and the CO emission shows velocity components delineating the edges of both objects (see Fig.~\ref{fits}a), supporting a scenario in which the two are genuinely interacting. It should be noted that all of the CO emission components show peak brightness temperatures that are unusually low for a GMC ($T_R\la5$ K), strongly suggesting that they belong to the same physical system and are not chance superpositions along the line of sight.

Approximating the GMC as a circular disk sandwiched between the two shells, with a diameter equal to the length of its bright ridge ($\sim100$ pc), we may estimate the average initial number density required to sweep up the molecular gas from the pre-shell medium. Taking the centres of expansion for the two shells as $(l,b)=(287.5^\circ,3.0^\circ)$ and $(290.1^\circ,0.2^\circ)$ \citep{rizzo98}, and the GMC mass as $M_{\mathrm{H}_2}=1.4\pm0.6\times10^5~\mathrm{M}_{\odot}$ we obtain an average initial H{\sc i} number density of $\sim10$ cm$^{-3}$. This implies that pre-existing dense gas would have been needed to form the present GMC, but tells us little about whether this material was predominantly atomic or already collected into dense molecular clouds. \citet{matsunagaphd} suggests that G288.5+1.5 is a young, pre-star-forming GMC, based on its apparent lack of massive star formation activity and unusually low density. Much current thought on molecular cloud and star formation favours a scenario in which the the entire process of cloud formation and dissipation occurs rapidly in around a crossing time ($\sim10^7$ yr), and star formation commences swiftly after the initial cloud is formed \citep{elmegreen00, hartmann01, maclow04}. If the GMC is indeed pre-star-forming, this would suggest an age of no more than a couple of Myr, and support a scenario in which its formation is either triggered or enhanced by the presence of the two shells.

Further investigation is now needed to explore these issues in more depth, and to place this somewhat speculative discussion on firmer physical and astronomical ground.

\section{Summary and Concluding Remarks}
\label{summary}

The Parkes SGPS H{\sc i} data have revealed the morphology of the Carina Flare supershell for the first time. The shell is well described by a partial ellipse centred on $(l,b)\approx(287^\circ,4^\circ)$, measuring $\sim230\times360$ pc, with its major axis inclined to the right of the vertical. The ellipse is broken at its top left, and a capped, high-latitude extension protrudes from the $\sim200$ pc gap, reaching $z\sim450$ pc. The expansion velocity and velocity centroid of the supershell are found to be $\sim10$ km s$^{-1}$ and $v_{lsr}\approx-17$ km s$^{-1}$, the latter adjusted slightly from F99's original estimate. H{\sc i} linewidths reach values as low as $4$-$6$ km s$^{-1}$, suggesting the presence of cold atomic gas within the walls. \par

The morphology of the supershell is well explained by a scenario in which it has broken at a height of $z\sim280$ pc and is blowing out of the Galactic Plane. It is not yet clear whether it is yet fully open to the Halo. However, it has already expanded to the disc--halo interface, and it is clear that the expansion has not yet stalled, suggesting that it is well on its way to becoming a fully-fledged Galactic chimney. 

The molecular clouds form co-moving parts of the atomic shell. The SGPS data reveal rich H{\sc i} substructure, and the CO and H{\sc i} components exhibit a variety of intriguing morphological relationships. We see evidence of molecular gas distributed along the inner rim of a thick, H{\sc i} filament; intense molecular emission delineating the shell's curved bottom rim,
and molecular clouds associated with the high-latitude extremes of the H{\sc i} shell. The Carina Flare supershell is unique in the extent and quality of the observed relationships between its atomic and molecular matter, and we suggest it may be unparalleled as a laboratory for investigating the effect of large-scale stellar feedback on the evolution and structure of the ISM; particularly the proposed role of supershells as molecular cloud producers. A simplistic consideration of time-scales suggests that the formation of molecular gas from the swept-up medium is viable, 
and we have briefly examined the case of the GMC G288.5+1.5, which we argue is an excellent candidate for triggered formation at the interface of a pair of converging supershells.

We have newly identified an ionized component of the supershell in the form of delicate, streamer-like filaments aligned with the proposed direction of blowout. These have projected lengths and widths of $50\sim100$ pc and $2\sim5$ pc respectively, and represent the ionized edges of parts of the shell ISM. A genuine association with the supershell is confirmed via their correlation with neutral gas features in the approaching limb, and their morphology suggests a relation to the blowout process. 

We have presented strong evidence to suggest that the Carina Flare supershell is associated with the Carina Spiral Arm feature, at a distance of $2.6\pm0.4$ kpc; consistent with the previous estimate of \citet{fukui99}. We have discussed its morphology in the context of large-scale horizontal density gradients, and speculate on the feasibility of a location towards the near edge of the Carina Arm.

The masses of the atomic and molecular ISM in the swept-up shell are estimated as $M_\mathrm{HI}\approx7\pm3\times10^5~\mathrm{M}_{\odot}$ and $M_{\mathrm{H}_2}\approx2.0\pm0.6\times10^5~\mathrm{M}_{\odot}$, respectively. The latter figure is almost 3 times larger than that of F99, who did not include the GMC G288.5+1.5 in their analysis. An average initial number density of $n_0\sim3$ cm$^{-3}$ would be required in order for the shell to have swept up all its mass from the ambient ISM.\par

The age and energetics of the supershell have been explored in a variety of ways. The kinetic energy of the expanding shell is estimated as $E_K\sim1\times10^{51}$, and a `dynamic age' of  $\tau_k\sim1\times10^7$ yr is derived by assuming a constant expansion velocity. Comparisons with simple analytical models yield formation energy and age estimates of $E\sim5\times10^{51}$ erg and $\tau\sim0.8\times10^7$ yr, respectively. However, the model assumptions are clearly not fulfilled and the estimates are unavoidably crude. Additionally, requirements for Galactic Plane blowout from numerical models suggest a mechanical luminosity input of at least a few $10^{37}$ erg s$^{-1}$ and an age of at least several $10^6$ yr (where these figures are not independent). Our formation energy estimates are around an order of magnitude smaller than those of F99, due to an error in their paper. \par  

We extend our warmest thanks to all the staff and students of Nagoya University who contributed to the observations utilised in this paper, and to the Las Campanas Observatory staff for their support and hospitality. The NANTEN project was based on a mutual agreement between Nagoya University and the Carnegie Institute of Washington, and its operation was made possible thanks to contributions from many companies and members of the Japanese public. We also gratefully acknowledge the Southern H-Alpha Sky Survey Atlas (SHASSA), which is supported by the National Science Foundation. J. Dawson receives financial support from a Japanese Government (MEXT) scholarship for postgraduate research studies.

\bibliography{mybibliography}

\begin{thebibliography}{60}
\expandafter\ifx\csname natexlab\endcsname\relax\def\natexlab#1{#1}\fi

\bibitem[{{Alvarez} {et~al.}(1990){Alvarez}, {May}, \& {Bronfman}}]{alvarez90}
{Alvarez} H., {May} J., {Bronfman} L., 1990, \apj, 348, 495

\bibitem[{{Bergin} {et~al.}(2004){Bergin}, {Hartmann}, {Raymond}, \&
  {Ballesteros-Paredes}}]{bergin04}
{Bergin} E.~A., {Hartmann} L.~W., {Raymond} J.~C., {Ballesteros-Paredes} J.,
  2004, \apj, 612, 921, arXiv:astro-ph/0405329

\bibitem[{{Boggs} {et~al.}(1992){Boggs}, {Byrd}, {Rogers}, \&
  {Schnabel}}]{boggs92}
{Boggs} P.~T., {Byrd} R.~H., {Rogers} J.~E., {Schnabel} R. B., 1992, {User's
  Reference Guide for ODRPACK Version 2.01 -- Software for Weighted Orthogonal
  Distance Regression}. National Institute of Standards and Technology NISTIR
  4834

\bibitem[{{Brand} \& {Blitz}(1993)}]{brand93}
{Brand} J., {Blitz} L., 1993, \aap, 275, 67

\bibitem[{{Breitschwerdt} \& {de Avillez}(2006)}]{breit06}
{Breitschwerdt} D., {de Avillez} M.~A., 2006, \aap, 452, L1, arXiv:astro-ph/0604162

\bibitem[{{Bruhweiler} {et~al.}(1980){Bruhweiler}, {Gull}, {Kafatos}, \&
  {Sofia}}]{bruh80}
{Bruhweiler} F.~C., {Gull} T.~R., {Kafatos} M., {Sofia} S., 1980, \apjl, 238,
  L27

\bibitem[{{Callaway} {et~al.}(2000){Callaway}, {Savage}, {Benjamin}, {Haffner},
  \& {Tufte}}]{callaway00}
{Callaway} M.~B., {Savage} B.~D., {Benjamin} R.~A., {Haffner} L.~M., {Tufte}
  S.~L., 2000, \apj, 532, 943

\bibitem[{{Chen} {et~al.}(2000){Chen}, {Chu}, {Gruendl}, \& {Points}}]{chen00}
{Chen} C.-H.~R., {Chu} Y.-H., {Gruendl} R.~A., {Points} S.~D., 2000, \aj, 119,
  1317

\bibitem[{{Clemens}(1985)}]{clemens85}
{Clemens} D.~P., 1985, \apj, 295, 422

\bibitem[{{Cordes} \& {Lazio}(2002)}]{cordes02}
{Cordes} J.~M., {Lazio} T.~J.~W., 2002, ArXiv Astrophysics e-prints, astro-ph/0207156

\bibitem[{{Dawson} {et~al.}(2007){Dawson}, {Kawamura}, {Mizuno}, {Onishi}, \&
  {Fukui}}]{dawson07}
{Dawson} J., {Kawamura} A., {Mizuno} N., {Onishi} T., {Fukui} Y., 2007, in IAU
  Symposium, Vol. 237, IAU Symposium, {Elmegreen} B.~G., {Palous} J., eds., pp.
  406--406

\bibitem[{{de Avillez} \& {Berry}(2001)}]{avillez01}
{de Avillez} M.~A., {Berry} D.~L., 2001, \mnras, 328, 708

\bibitem[{{de Avillez} \& {Breitschwerdt}(2005)}]{avillez05}
{de Avillez} M.~A., {Breitschwerdt} D., 2005, \aap, 436, 585, arXiv:astro-ph/0502327

\bibitem[{{Dickey} \& {Lockman}(1990)}]{dickey90}
{Dickey} J.~M., {Lockman} F.~J., 1990, \araa, 28, 215

\bibitem[{{Dickey} {et~al.}(2001){Dickey}, {McClure-Griffiths},
  {Stanimirovi{\'c}}, {Gaensler}, \& {Green}}]{dickey01}
{Dickey} J.~M., {McClure-Griffiths} N.~M., {Stanimirovi{\'c}} S., {Gaensler}
  B.~M., {Green} A.~J., 2001, \apj, 561, 264, arXiv:astro-ph/0107604

\bibitem[{{Ehlerov{\'a}} \& {Palou{\v s}}(2005)}]{ehlerova05}
{Ehlerov{\'a}} S., {Palou{\v s}} J., 2005, \aap, 437, 101, arXiv:astro-ph/0503443

\bibitem[{{Elmegreen}(1998)}]{elmegreen98}
{Elmegreen} B.~G., 1998, in Astronomical Society of the Pacific Conference
  Series, Vol. 148, Origins, {Woodward} C.~E., {Shull} J.~M., {Thronson} Jr.
  H.~A., eds., pp. 150--+

\bibitem[{{Elmegreen}(2000)}]{elmegreen00}
---, 2000, \apj, 530, 277, arXiv:astro-ph/9911172

\bibitem[{{Fich} {et~al.}(1989){Fich}, {Blitz}, \& {Stark}}]{fich89}
{Fich} M., {Blitz} L., {Stark} A.~A., 1989, \apj, 342, 272

\bibitem[{{Fukui} {et~al.}(1999){Fukui}, {Onishi}, {Abe}, {Kawamura},
  {Tachihara}, {Yamaguchi}, {Mizuno}, \& {Ogawa}}]{fukui99}
{Fukui} Y., {Onishi} T., {Abe} R., {Kawamura} A., {Tachihara} K., {Yamaguchi}
  R., {Mizuno} A., {Ogawa} H., 1999, \pasj, 51, 751

\bibitem[{{Garcia}(1994)}]{garcia94}
{Garcia} B., 1994, \apj, 436, 705

\bibitem[{{Gaustad} {et~al.}(2001){Gaustad}, {McCullough}, {Rosing}, \& {Van
  Buren}}]{gaustad01}
{Gaustad} J.~E., {McCullough} P.~R., {Rosing} W., {Van Buren} D., 2001, \pasp,
  113, 1326, arXiv:astro-ph/0108518

\bibitem[{{Georgelin} {et~al.}(2000){Georgelin}, {Russeil}, {Amram},
  {Georgelin}, {Marcelin}, {Parker}, \& {Viale}}]{georg00}
{Georgelin} Y.~M., {Russeil} D., {Amram} P., {Georgelin} Y.~P., {Marcelin} M.,
  {Parker} Q.~A., {Viale} A., 2000, \aap, 357, 308

\bibitem[{{Glover} \& {Mac Low}(2007)}]{glover07}
{Glover} S.~C.~O., {Mac Low} M.-M., 2007, \apj, 659, 1317, arXiv:astro-ph/0605121

\bibitem[{{Grabelsky} {et~al.}(1987){Grabelsky}, {Cohen}, {Bronfman},
  {Thaddeus}, \& {May}}]{grab87}
{Grabelsky} D.~A., {Cohen} R.~S., {Bronfman} L., {Thaddeus} P., {May} J., 1987,
  \apj, 315, 122

\bibitem[{{Green}(1993)}]{green93}
{Green} D.~A., 1993, \mnras, 262, 327

\bibitem[{{Hartmann} {et~al.}(2001){Hartmann}, {Ballesteros-Paredes}, \&
  {Bergin}}]{hartmann01}
{Hartmann} L., {Ballesteros-Paredes} J., {Bergin} E.~A., 2001, \apj, 562, 852, arXiv:astro-ph/0108023

\bibitem[{{Heiles}(1979)}]{heiles79}
{Heiles} C., 1979, \apj, 229, 533

\bibitem[{{Heiles}(1984)}]{heiles84}
---, 1984, \apjs, 55, 585

\bibitem[{{Hunter} {et~al.}(1997){Hunter}, {Bertsch}, {Catelli}, {Dame},
  {Digel}, {Dingus}, {Esposito}, {Fichtel}, {Hartman}, {Kanbach}, {Kniffen},
  {Lin}, {Mayer-Hasselwander}, {Michelson}, {von Montigny}, {Mukherjee},
  {Nolan}, {Schneid}, {Sreekumar}, {Thaddeus}, \& {Thompson}}]{hunter97}
{Hunter} S.~D., {Bertsch} D.~L., {Catelli} J.~R., {Dame} T.~M., {Digel} S.~W.,
  {Dingus} B.~L., {Esposito} J.~A., {Fichtel} C.~E., {Hartman} R.~C., {Kanbach}
  G., {Kniffen} D.~A., {Lin} Y.~C., {Mayer-Hasselwander} H.~A., {Michelson}
  P.~F., {von Montigny} C., {Mukherjee} R., {Nolan} P.~L., {Schneid} E.,
  {Sreekumar} P., {Thaddeus} P., {Thompson} D.~J., 1997, \apj, 481, 205

\bibitem[{{Igumenshchev} {et~al.}(1990){Igumenshchev}, {Shustov}, \&
  {Tutukov}}]{igumen90}
{Igumenshchev} I.~V., {Shustov} B.~M., {Tutukov} A.~V., 1990, \aap, 234, 396

\bibitem[{{Kerr} {et~al.}(1986){Kerr}, {Bowers}, {Jackson}, \& {Kerr}}]{kerr86}
{Kerr} F.~J., {Bowers} P.~F., {Jackson} P.~D., {Kerr} M., 1986, \aaps, 66, 373

\bibitem[{{Leitch} \& {Vasisht}(1998)}]{leitch98}
{Leitch} E.~M., {Vasisht} G., 1998, New Astronomy, 3, 51, arXiv:astro-ph/9802174

\bibitem[{{Lockman}(1984)}]{lockman84}
{Lockman} F.~J., 1984, \apj, 283, 90

\bibitem[{{Mac Low} \& {Klessen}(2004)}]{maclow04}
{Mac Low} M.-M., {Klessen} R.~S., 2004, Reviews of Modern Physics, 76, 125, arXiv:astro-ph/0301093

\bibitem[{{Mac Low} \& {McCray}(1988)}]{maclow88}
{Mac Low} M.-M., {McCray} R., 1988, \apj, 324, 776

\bibitem[{{Mac Low} {et~al.}(1989){Mac Low}, {McCray}, \& {Norman}}]{maclow89}
{Mac Low} M.-M., {McCray} R., {Norman} M.~L., 1989, \apj, 337, 141

\bibitem[{{Malhotra}(1994)}]{malhotra94a}
{Malhotra} S., 1994, \apj, 433, 687, arXiv:astro-ph/9404028

\bibitem[{{Matsunaga}(2002)}]{matsunagaphd}
{Matsunaga} K., 2002, PhD thesis, Nagoya University

\bibitem[{{Matsunaga} {et~al.}(2001){Matsunaga}, {Mizuno}, {Moriguchi},
  {Onishi}, {Mizuno}, \& {Fukui}}]{matsunaga01}
{Matsunaga} K., {Mizuno} N., {Moriguchi} Y., {Onishi} T., {Mizuno} A., {Fukui}
  Y., 2001, \pasj, 53, 1003

\bibitem[{{McClure-Griffiths} {et~al.}(2002){McClure-Griffiths}, {Dickey},
  {Gaensler}, \& {Green}}]{mcclg02}
{McClure-Griffiths} N.~M., {Dickey} J.~M., {Gaensler} B.~M., {Green} A.~J.,
  2002, \apj, 578, 176, arXiv:astro-ph/0206358

\bibitem[{{McClure-Griffiths} {et~al.}(2003){McClure-Griffiths}, {Dickey},
  {Gaensler}, \& {Green}}]{mcclg03}
---, 2003, \apj, 594, 833

\bibitem[{{McClure-Griffiths} {et~al.}(2005){McClure-Griffiths}, {Dickey},
  {Gaensler}, {Green}, {Haverkorn}, \& {Strasser}}]{mcclg05}
{McClure-Griffiths} N.~M., {Dickey} J.~M., {Gaensler} B.~M., {Green} A.~J.,
  {Haverkorn} M., {Strasser} S., 2005, \apjs, 158, 178, arXiv:astro-ph/0503134

\bibitem[{{McClure-Griffiths} {et~al.}(2006){McClure-Griffiths}, {Ford},
  {Pisano}, {Gibson}, {Staveley-Smith}, {Calabretta}, {Dedes}, \&
  {Kalberla}}]{mcclg06}
{McClure-Griffiths} N.~M., {Ford} A., {Pisano} D.~J., {Gibson} B.~K.,
  {Staveley-Smith} L., {Calabretta} M.~R., {Dedes} L., {Kalberla} P.~M.~W.,
  2006, \apj, 638, 196, arXiv:astro-ph/0510304

\bibitem[{{McClure-Griffiths} {et~al.}(2001){McClure-Griffiths}, {Green},
  {Dickey}, {Gaensler}, {Haynes}, \& {Wieringa}}]{mcclg01}
{McClure-Griffiths} N.~M., {Green} A.~J., {Dickey} J.~M., {Gaensler} B.~M.,
  {Haynes} R.~F., {Wieringa} M.~H., 2001, \apj, 551, 394, arXiv:astro-ph/0012302

\bibitem[{{McCray} \& {Kafatos}(1987)}]{mccray87}
{McCray} R., {Kafatos} M., 1987, \apj, 317, 190

\bibitem[{{Moriguchi} {et~al.}(2002){Moriguchi}, {Onishi}, {Mizuno}, \&
  {Fukui}}]{moriguchi02}
{Moriguchi} Y., {Onishi} T., {Mizuno} A., {Fukui} Y., 2002, in 8th
  Asian-Pacific Regional Meeting, Volume II, {Ikeuchi} S., {Hearnshaw} J.,
  {Hanawa} T., eds., pp. 173--174

\bibitem[{{Norman} \& {Ikeuchi}(1989)}]{norman89}
{Norman} C.~A., {Ikeuchi} S., 1989, \apj, 345, 372

\bibitem[{{Normandeau} {et~al.}(1996){Normandeau}, {Taylor}, \&
  {Dewdney}}]{ndeau96}
{Normandeau} M., {Taylor} A.~R., {Dewdney} P.~E., 1996, \nat, 380, 687

\bibitem[{{Ogawa} {et~al.}(1990){Ogawa}, {Mizuno}, {Ishikawa}, {Fukui}, \&
  {Hoko}}]{ogawa90}
{Ogawa} H., {Mizuno} A., {Ishikawa} H., {Fukui} Y., {Hoko} H., 1990,
  International Journal of Infrared and Millimeter Waves, 11, 717

\bibitem[{{Rizzo} \& {Arnal}(1998)}]{rizzo98}
{Rizzo} J.~R., {Arnal} E.~M., 1998, \aap, 332, 1025

\bibitem[{{Russeil}(2003)}]{russeil03}
{Russeil} D., 2003, \aap, 397, 133

\bibitem[{{Taylor} \& {Cordes}(1993)}]{taylor93}
{Taylor} J.~H., {Cordes} J.~M., 1993, \apj, 411, 674

\bibitem[{{Tenorio-Tagle} {et~al.}(1990){Tenorio-Tagle}, {Rozyczka}, \&
  {Bodenheimer}}]{ttagle90}
{Tenorio-Tagle} G., {Rozyczka} M., {Bodenheimer} P., 1990, \aap, 237, 207

\bibitem[{{Tomisaka}(1992)}]{tomisaka92}
{Tomisaka} K., 1992, \pasj, 44, 177

\bibitem[{{Tomisaka}(1998)}]{tomisaka98}
---, 1998, \mnras, 298, 797, arXiv:astro-ph/9804029

\bibitem[{{Tomisaka} {et~al.}(1981){Tomisaka}, {Habe}, \&
  {Ikeuchi}}]{tomisaka81}
{Tomisaka} K., {Habe} A., {Ikeuchi} S., 1981, \apss, 78, 273

\bibitem[{{Tomisaka} \& {Ikeuchi}(1986)}]{tomisaka86}
{Tomisaka} K., {Ikeuchi} S., 1986, \pasj, 38, 697

\bibitem[{{Weaver} {et~al.}(1977){Weaver}, {McCray}, {Castor}, {Shapiro}, \&
  {Moore}}]{weaver77}
{Weaver} R., {McCray} R., {Castor} J., {Shapiro} P., {Moore} R., 1977, \apj,
  218, 377

\bibitem[{{Yamaguchi} {et~al.}(1999){Yamaguchi}, {Mizuno}, {Moriguchi},
  {Yonekura}, {Mizuno}, \& {Fukui}}]{yamaguchi99}
{Yamaguchi} N., {Mizuno} N., {Moriguchi} Y., {Yonekura} Y., {Mizuno} A.,
  {Fukui} Y., 1999, \pasj, 51, 765

\end{thebibliography}

\appendix
\section{Fitting to the $l$-$v_{lsr}$ Cavity}
\label{fitting}

A well-behaved expanding shell appears in any given plane of a 3D datacube as a low brightness cavity bordered by a bright rim. Reliable measurements of the size and location of this void provide a means of quantifying trends in a shell's appearance through a datacube, and a robust basis from which to discuss spatial and velocity structure. However, such measurements are not usually attempted, perhaps because of the `messiness' of many H{\sc i} supershells. We have developed a simple, semi-objective method of fitting to cavities in the longitude-velocity plane. Although something of an exploratory exercise, it is outlined here because of its potential to provide standardised, automated measurements across large samples of objects.\par 

\begin{figure*} 
\includegraphics{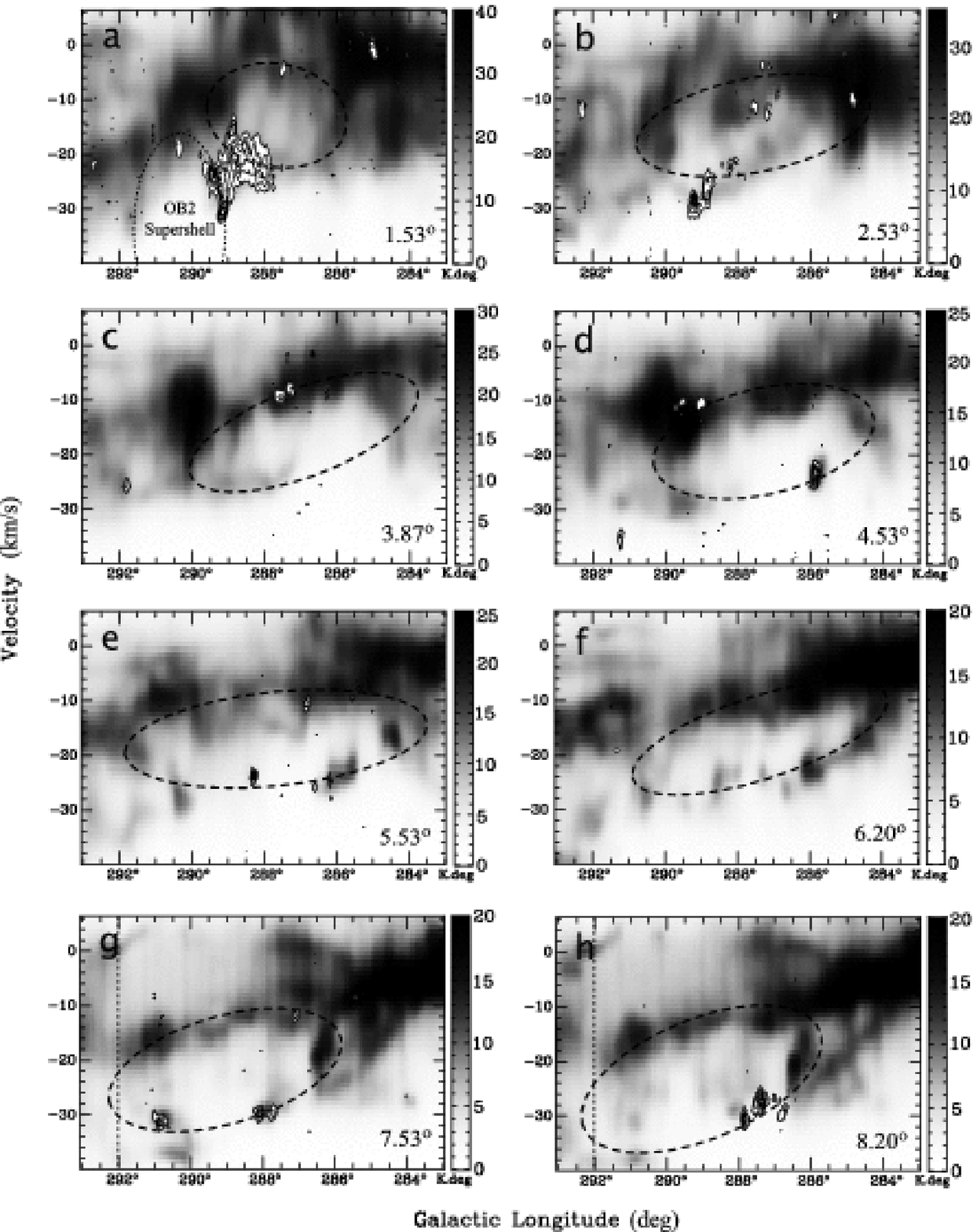}   
\caption{Selected $l$-$v_{lsr}$ slices, integrated over latitude intervals of $0.33^\circ$. The central latitude of each slice is labelled. The grayscale image is H{\sc i} and contours are $^{12}$CO(J=1-0) at contour levels of 0.1 + 0.2 K.deg. The dashed ellipses are orthogonal distance regression fits to sets of points characterising the supershell. Thin dotted lines mark the boundary of the NANTEN $^{12}$CO observations. The approximate location of the Carina OB2 supershell is marked by eye on the $b=1.53^{\circ}$ slice.}
\label{fits}
\end{figure*}

The basic method may be summarised as follows. We divide the datacube into $l$-$v_{lsr}$ slices, each integrated over a latitude interval of 20 arcmin (just over one spatial resolution element). A sample of smoothed 1D H{\sc i} emission profiles are taken along radial cuts through a point in the centre of the cavity, and the first significant emission peaks after the central trough are identified as the bright rim of the supershell. The molecular data are incorporated by locating islands of CO emission and including their peaks in the point set. A robust orthogonal distance regression package, \citep[ODRPACK,][]{boggs92} is then used to fit ellipses to the collection of points, where the ellipses are allowed to rotate freely within the plane as an approximate method of adjusting to the systematic slope that arises due to Galactic rotation. The fitting success rate is around $\sim80$ per cent, with the remaining $\sim20$ per cent returning point sets for which the present fitting routine will not converge to a realistic result. Because we are fitting to multiple 2D slices, however, this irregular fifth of the sample may be discarded without appreciable loss of information about the large-scale trends in the shell's appearance. Fig.~\ref{fits} shows a sample of the most robust fits in each $1^\circ$ latitude interval, overlaid on the CO and H{\sc i} data.\par 

Uncertainties are crudely estimated by varying the number, orientation and centre point of the 1D emission profiles, and relaxing and tightening the criteria for the removal of outliers. The velocity parameters are generally most robust, with $v_0$ (the velocity of the ellipse centre) varying by $\la1$ km s$^{-1}$ and the velocity widths varying by $\la2$ km s$^{-1}$. A slight systematic overestimation in longitude width occurs in the present implementation, which arises from our decision to set one $v_{lsr}$ pixel `spatially' equal to one $l$ pixel. This causes the $l$-$v_{lsr}$ holes to take on an elongated appearance, meaning that the ODR routine can sacrifice spatial accuracy relatively inexpensively in order to minimize deviations in the velocity dimension.\par   

Nevertheless, the routine is successful in that it recovers and parametrizes variations in the width, location and velocity structure of the shell in a way that is consistent with its appearance in the $l$-$b$ and $l$-$v_{lsr}$ planes. It therefore has the potential to be developed further into a fully automated and consistent method of obtaining quantitative information about the spatio-velocity structure of expanding shells.\par 

In the case of the Carina Flare supershell, the scientific usefulness of the fits is mitigated by the presence of the large non-local emission component. This may act to distort the apparent shape of the shell in the $l$-$v_{lsr}$ plane, reducing the reliability of the fitted parameters as genuine descriptors of its structure. The severity of the contribution is also latitude dependent, which could potentially produce false latitude-dependent trends in the fits. We must therefore be very cautious in interpreting any apparent features. Nevertheless, it is interesting to note that the highest latitude fits in Fig.~\ref{fits} ($b=7.5^{\circ}, 8.2^{\circ}$) show what appears to be a genuine offset towards negative velocities. These slices correspond to the high-latitude extension of the supershell, and this is suggestive of preferential forward expansion in the region.

\end{document}